# Plate-like precipitate effects on plasticity of Al-Cu micro-pillar: {100}-interfacial slip

Peng Zhang[a*], Jian-Jun Bian[a*], Chong Yang[a], Jin-Yu Zhang[a], Gang Liu[a], Jérôme Weiss[b], Jun Sun[a]

[a]*State Key Laboratory for Mechanical Behavior of Materials, Xi'an Jiaotong University, Xi'an, 710049, China*
[b]*IsTerre, CNRS/Université Grenoble Alpes, 38401 Grenoble, France*

**Abstract** ─ In this paper, we study the effects of $\theta'$-Al$_2$Cu plate-like precipitates on the plasticity of Al-Cu micro-pillars, with a sample size allowing the precipitates to cross the entire micro-pillar. {100}-slip traces are identified for the first time in Al and Al alloys at room temperature. We investigate the underlying mechanisms of this unusual {100}-slip, and show that it operates along the coherent $\theta'$-Al$_2$Cu precipitate/$\alpha$-Al matrix interface. A combination of molecular dynamics simulations and stress analysis indicates that screw dislocations can cross-slip from the {111} plane onto the {100} $\theta'$-Al$_2$Cu/$\alpha$-Al interface, then move on it through a kink-pair mechanism, providing a reasonable explanation to the observed {100}-slips. The roles of the $\theta'$-Al$_2$Cu precipitate/$\alpha$-Al matrix interface on the properties of interfacial dislocations are studied within the Peierls-Nabarro framework, showing that the interface can stabilize the {100} screw dislocations from the spreading of the core, and increases the Peierls stress. These results improve our understanding of the mechanical behavior of Al-Cu micro-pillars at room temperature, and imply an enhanced role of interfacial slip in Al-Cu based alloys at elevated temperature in consideration of the underlying kink-pair mechanism.

*These authors contributed equally to this work.

Correspondence and requests for materials should be addressed to lgsammer@xjtu.edu.cn (GL), or jerome.weiss@univ-grenoble-alpes.fr (JW), or junsun@xjtu.edu.cn (JS)

# 1. Introduction

Aluminum alloys have been widely used in the automotive and aviation industries as a structural material due to their excellent combination of high strength and light weight [1]. The binary Al-Cu alloy is one of the most important model systems in the study of Al alloys, as it serves as a basis for a wide range of age-hardened Al alloys [2]. One of the most effective strengthening precipitates in Al-Cu systems is the $\theta'$-Al$_2$Cu plate-like intermetallic, the diameter of which can be artificially tailored from several ten nanometers to micrometers by heat treatment or a micro-alloying methodology [3-6]. The significant strength improvement in Al-Cu alloy was believed to stem from the Orowan mechanism, based on the conventional knowledge that the $\theta'$-Al$_2$Cu phase is shear-resistant to dislocations [3, 7-9]. However, in severely deformed Al-Cu alloys [5, 10] and Al-Cu micro-pillars [11], a shearing of $\theta'$-Al$_2$Cu precipitates has been observed, giving an indication that the underlying precipitate-dislocation interactions are more complex than the classical Orowan mechanism. Indeed, a recent micro-mechanical testing combined with transmission X-ray microscopy supports a transition of strengthening modes in Al-Cu alloy, evolving from dislocation bypassing to dislocation accumulation as increasing the precipitate diameter [12].

To further explore other possible plastic mechanisms related to $\theta'$-Al$_2$Cu precipitates in Al-Cu alloy, we employ the micro-pillar compression methodology in this paper. Comparing with previous investigations [12-14], the particularity of our samples is that the average diameter of the precipitates is commensurate with the sample diameter. In this case, the plate-like precipitates have a large possibility to cut the entire micro-pillar, thus suppressing the Orowan mechanism and allowing us to explore other novel dislocation-precipitate interactions through this special configuration. Indeed, we observe non-trivial slip traces along {100} planes. In face-centered cubic (FCC) metallic crystals deformed at room temperature or below, dislocation gliding along {111} octahedral planes is generally considered as the only significant slip mechanism, owing to numerous equivalent slip systems and a low Peierls stress. According to our knowledge, this is the first time that {100}-slip traces are reported at *room* temperature in an Al alloy.

On the other hand, in the last century, several slip-lines analyses in Al crystals reported an enhanced role of non-octahedral slips along {100} planes, as well as {110}, {112} and {113} planes, upon increasing temperature [15-17]. Carrard and Martin [18, 19] revealed that in [112]-oriented pure Al single crystals, {100}-slip is present from a temperature of 180 ºC, accompanied by frequent cross-slips between {100} and {111} planes, and operates alone at 400 ºC and above. It was shown that this {100}-slip comes from {100} dislocations multiplying on their own slip planes, rather than from dislocations cross-slipping from intersecting {111} planes [18, 19]. Moreover, slip along non-close-packed planes proceed through a kink-pair mechanism, as a result of the dissociation of screw dislocations on the intersecting close-packed plane [20]. In this case, gliding of the dissociated screw dislocations should be facilitated by thermal fluctuations, forming edge-kinks with a high mobility on non-close-packed planes [19, 20].

The activation of non-octahedral slips provides an additional degree of freedom for plastic deformation, thereby playing an important role in the mechanical behavior at elevated temperature, essentially in terms of strain hardening [21] and of texture evolution during severe plastic deformation [22]. For instance, both experiments on hot-rolling [23] and numerical simulations [22] have demonstrated that the {110} and {100} slips can stabilize the ideal orientation of texture in FCC crystals, and hence should affect the subsequent recrystallization microstructures, as well as their mechanical properties. Despite their potential consequences in terms of mechanical behavior, compared with non-octahedral slips operating only at *elevated* temperature, non-octahedral slips operating in Al alloys at *room* temperature, which must be associated with precipitates, remain unknown. Based on this consideration, besides the observation of novel {100}-slip, we also analyze here the role of $\theta'$-$Al_2Cu$ precipitates on the activation of {100}-slip by using a combination of computational simulations and theoretical analyses, with the aim to provide fundamental insights into this non-trivial slip mode in Al-Cu alloys. Our results are expected to improve our understanding of the plasticity in Al-Cu based alloys, with the potential for optimizing the mechanical properties from interfacial engineering.

## 2. Experimental and simulation methods

*2.1 Experiments*

The Al-4.0wt.% Cu alloy was melted and cast in a stream argon, by using 99.99 wt.% pure Al and Al-50 wt.% Cu master-alloy. The Al-Cu alloy was homogenized at 723 K for 24 h to eliminate composition segregation, and then solutionized at 793 K for 4 h followed by a water quench. This alloy was further aged at 523 K for 8 h to form $\theta'$-Al$_2$Cu precipitates, which were characterized by standard transmission electron microscopy (TEM) and high resolution transmission electron microscopy (HRTEM).

The micro-pillars on the electro-polished surface of a [110]-grain of the aged Al-Cu alloy were fabricated by using a Ga-operated focus iron beam (FIB). The diameter of micro-pillars was 1000 nm, with an aspect ratio of ~ 3:1. A nano-indentation system (Hysitron Ti 950) was used to compress the micro-pillars at room temperature under a displacement-controlled mode, with a strain rate of ~$2 \times 10^{-4}$ s$^{-1}$ up to 18% engineering strain.

*2.2 Molecular dynamics (MD) simulations*

We performed classical molecular dynamics simulations by using LAMMPS developed by Sandia National Laboratories [24]. 3D atomic models with dimensions $40 \times 30 \times 20$ nm$^3$ are used in the present study. Atomic interactions are described by the embedded-atom-method (EAM) which has been widely used for the Al-Cu metallic system [25]. Time evolution of the whole atomic system, with a timestep of 1.0 fs, is within the framework of microcanonical ensembles (NVE). The Berendsen thermostat is adopted to adjust temperature [26]. Initial dislocations (edge or screw type) are created near the free surface by artificially displacing atoms according to the displacement field of a dislocation [27].

*2.3 Ab-initio calculations*

To understand the atomistic mechanisms of {100}-slip near or at the precipitate/matrix interface, and to estimate the corresponding Peierls stress, *ab-initio* calculations were performed by using the plane-wave pseudopotential method implemented in Vienna ab initio Simulation Package (VASP). The simulated atomic super-cell, with periodic boundary conditions, was ~$5.73 \times 5.73 \times 32.40$ Å$^3$ in size, which consisted of 16 atom layers in the direction perpendicular to the slip plane. The maximum K-point grid for Brillouin zone integration was $10 \times 10 \times 1$. The generalized stacking fault (GSF) structure was created by rigidly moving one half of the crystal over the other half along the slip direction, and then relaxing atom coordinate components perpendicular to the slip plane. By computing the energy variations between the slipped and the initial atomic configurations, GSF energy curves were generated as a function of slip distance, based on pseudo-potentials together with a local-density approximation for the exchange and correlation energy [28]. We then combined the GSF energy curves with the classical Peierls-Nabarro (P-N) model [29, 30] by using the numerical methodology detailed in [31], giving both the core structure of the dislocation and the corresponding Peierls stress.

## 3. Results and discussions

In section 3.1, we present experimental evidences of {100}-interfacial slip and clarify that it is an intrinsic property of Al-Cu micro-pillars, rather than the extrinsic behavior resulting from the Ga injection during FIB fabrication. The origin of this {100}-interfacial slip is studied in section 3.2 through the MD simulations, and the associated atomistic mechanisms are detailed in section 3.3. In section 3.4, we compare the Peierls stress of interfacial slip and precipitate shearing, and confirm that the mechanisms observed in our MD simulations can indeed operate at the experimentally observed stresses in micro-pillars.

*3.1 Experimental evidences for {100}-slip*

After heat treatment, a large number of $\theta'$-$Al_2Cu$ precipitates are formed on the $\{100\}_\alpha$ plane of the Al-Cu alloy (TEM images in Fig. 1a and b). The $\theta'$-$Al_2Cu$ precipitates have a plate-shaped morphology with coherent $(100)_{\theta'}\|(100)_{\alpha\text{-Al}}$ interfaces along the broad face and semi-coherent interfaces around the rim of the plates [6]. The precipitate diameter can be directly measured from the TEM images, showing a broad size distribution (Fig. 1a, inset). However, TEM measurements inevitably exaggerate this scattering to some extent for stereological reasons, as the TEM foils, with a thickness (measured as ~ 70 nm) much smaller than the precipitate diameter, truncate the precipitates at a random position. Similarly, the real average precipitate diameter should be larger than the measured one, and can be corrected by $d_d = 2/\pi\left(d_m - t_f + \sqrt{(d_m - t_f)^2 + 4\pi d_m t_f}\right)$ [3], where $d_m$ and $d_d$ are measured and real mean diameters, respectively, and $t_f$ is the TEM foil thickness.

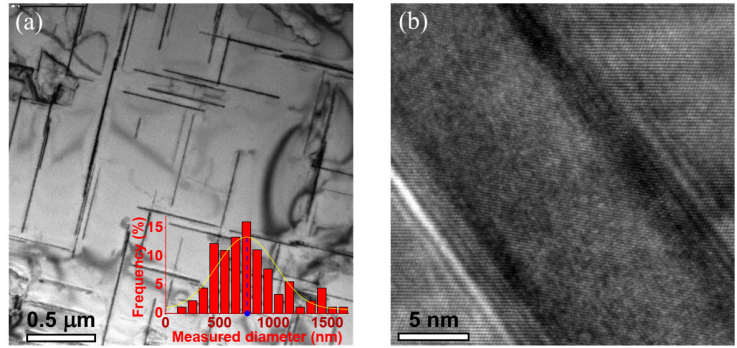

**Figure 1. Precipitates in the aged Al-Cu alloy. (a)** TEM image of the aged Al-Cu alloy, taken along the {100}-direction. The inset gives the distribution of measured precipitate diameters. **(b)** HRTEM image of $\theta'$-Al$_2$Cu precipitate, showing a thickness of ~ 15 nm.

The corrected average diameter $d_d$ is ~ 978 nm, very close to the diameter of micro-pillars (1000 nm, see Fig. 2a). Consequently, the plate-like precipitates have a large possibility to penetrate the micro-pillar. This unique structure rules out the classical Orowan mechanism, due to the absence of space for the dislocation to bow-out, hence allowing to study novel deformation mechanisms of Al-Cu alloys. Indeed, we identify some slip-traces clearly not associated with {111}-slip on the surface of deformed micro-pillar (Fig. 2b). A simple trace-shape analysis along several directions indicates that they could be a hint for {100}-slip (Fig. 2b, c and d). Owing to the fact that the intersection lines between the plate-like precipitates and the micro-pillar surface can be recognized before testing from the composition contrast (Fig. 2a), we found that these {100} slip-traces overlap with the intersection lines (see the arrows in Fig. 2a and b), implying dislocation glide on a {100} plane at or very close to the $\theta'$-Al$_2$Cu precipitate/$\alpha$-Al matrix coherent interface. As mentioned above, some previous works have revealed the operation of {100}-slip in pure Al at temperatures larger than 180 °C [16, 19, 20]. However, the underlying mechanisms should be different here as the {100}-slips occurred exactly at or very close to the precipitate/matrix interface, rather than in the Al matrix. Thus, the effects of this interface should be taken into consideration.

Some studies showed that liquid Ga can rapidly penetrate high-energy grain boundaries (GB) under stress, developing two layers of Ga-Al interfaces that lower the GB energy [32, 33]. Therefore, one may wonder if the Ga atoms injected on the micro-pillar surface during the focus ion beam (FIB) fabrication [34] might diffuse along the coherent {100}-interface, forming easy glide Ga-rich layers. To test this speculation, we extracted a longitudinal section thin foil from a micro-pillar fabricated by FIB, and then examined the Ga distribution on the penetrating plate-like $\theta'$-Al$_2$Cu precipitate. Note that the longitudinal section foil was also fabricated by FIB. The high-angle annular dark field (HAADF) image (Fig. 3a) reveals a straight $\theta'$-Al$_2$Cu precipitate inside the micro-pillar. However, the corresponding energy dispersive spectroscopy (EDS) Ga-mapping on Fig. 3b shows a uniform distribution of Ga, rather than a Ga-enriched layer at the interface. This uniformity on the entire section also indicates that the coherent $\theta'$-Al$_2$Cu/$\alpha$-Al interface is not a site of Ga ions segregation, even during the secondary Ga implantation process (FIB fabrication of the longitudinal section foil of the micro-pillar), in strong contrast with the severe Ga segregation at grain boundaries and incoherent interfaces in FIB milled Al thin foils [35]. In addition, a perfectly coherent interface can be observed by HRTEM (Fig. 3c and d), without any heterogeneous phase or additional Ga-related atom layer. In view of these observations, a possible Ga-layer slipping mechanism is unlikely, and the novel {100}-slip appears as an intrinsic property associated with the $\theta'$-Al$_2$Cu precipitate in Al-Cu alloys.

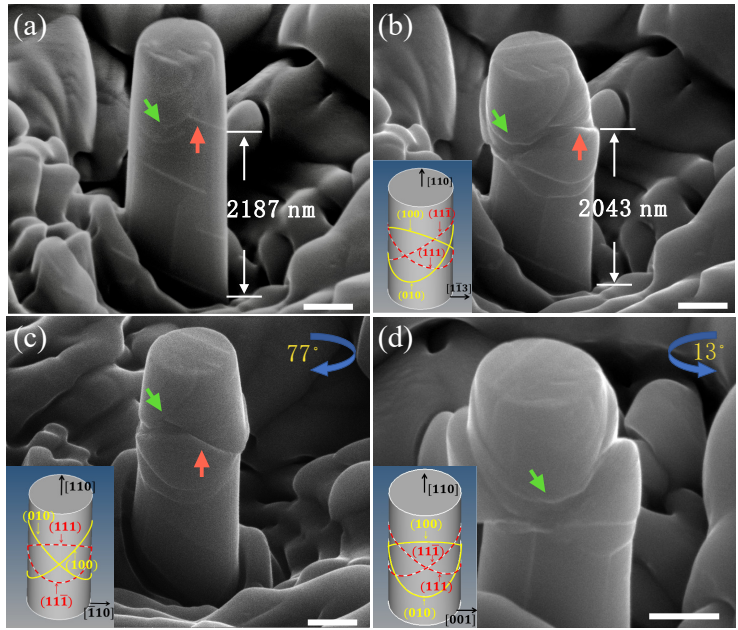

**Figure 2. Slip trace analysis.** (a) SEM image of an undeformed micro-pillar, where the traces of intersections between the plate-like precipitates and the surface of the micro-pillar can be identified from the composition contrast. (b) The deformed counterpart observed along the same direction as in (a). Non-octahedral slips occurring at the position of plate-like precipitates are marked by arrows. (c) and (d) The same micro-pillar observed along another

direction. The direction of rotation is given at the top right corner. The insets in (b), (c) and (d) show the shapes of slip traces along the corresponding observed directions. All the scale bars represent 500 nm.

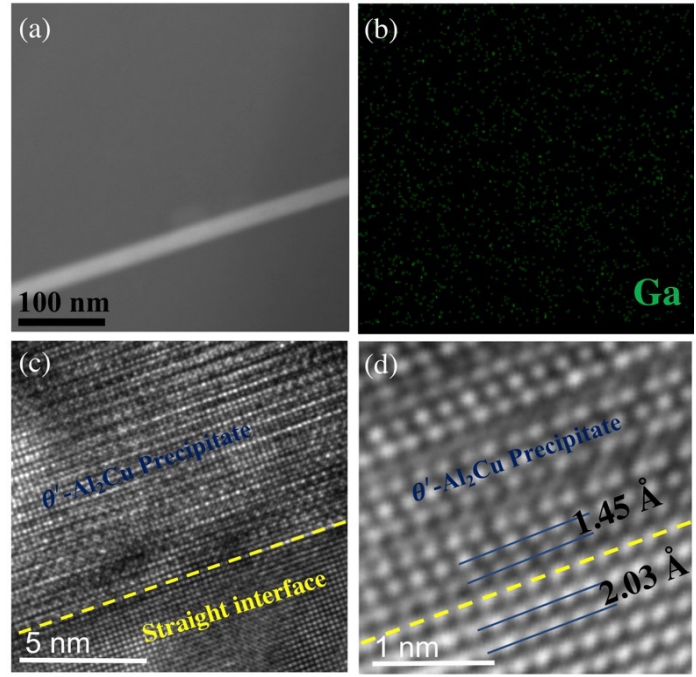

**Figure 3. Characterization of a $\theta'$-Al$_2$Cu precipitate penetrating the micro-pillars. (a)** and **(b)** are a HAADF image and its corresponding EDS Ga-mapping respectively. A uniform distribution of Ga atoms can be identified. **(c)** is a HRTEM image showing a sharp and straight interface between precipitate and Al matrix. **(d)** is a HRTEM image showing the interfacial atomic structure of (c).

*3.2 Interactions between dislocations and penetrating precipitates*

In view of the observations that the {100}-slips operate exactly along the plate-like precipitate, one may wonder about dislocation motion at the $\theta'$-Al$_2$Cu/$\alpha$-Al interface. Indeed, in metallic multi-layers, MD simulations showed that interfacial slip can result from dislocations nucleating in the regions of the interface with a large misfit strain [36, 37]. However, in our case, the $\theta'$-Al$_2$Cu/$\alpha$-Al interface is highly coherent (Fig. 3d), and thus should result in a nucleation stress much larger than that associated with semi-coherent or non-coherent interfaces. In this case, dislocation *nucleation* at the interface appears unlikely, and we speculate instead if dislocations *multiplying* from sources within the Al matrix could lead to {100}-slip when interacting with the interface. Consequently, we first study, from MD simulations, the interactions between matrix dislocations and penetrating $\theta'$-Al$_2$Cu precipitates to explore how these interactions can be linked to the non-trivial slip traces observed in the micro-pillars.

MD simulations of a full-sized micro-pillar lie beyond current computational power. Simulating a smaller pillar as a whole, in the nm-range, would imply large external stresses as the result of size effects on strength [38, 39]. This may trigger specific mechanisms unrealistic under the experimental stress levels. To avoid these drawbacks, we constructed a simple atomistic model in representing a local box inside the micro-pillar (see e.g. the box in Fig. 4a), and focussed only on the interaction mechanisms between dislocations and the precipitate. The stress requirements for the operation of the mechanisms observed in MD simulations will be discussed in section 3.4 from theoretical calculations. The atomic model contains a $\theta'$-Al$_2$Cu precipitate cutting through the Al matrix (Fig. 4b). Periodic boundary conditions are applied in the y-direction, while it is kept free in the x-direction. A constant displacement rate is imposed on the upper and lower surfaces, along the direction of Burgers vector. The simulation temperature is 298 K, identical with our experiments. The orientation of the tested micro-pillars was [110], so there are two equally easy slip planes, (111) and (11$\bar{1}$), with Burgers vectors $a/2[10\bar{1}]$ and $a/2[01\bar{1}]$, $a/2[101]$ and $a/2[011]$, respectively, where $a$ is the lattice constant of Al. In view of the orientation relation between these easy slip planes and the broad-plane of the precipitate, the dislocations depositing at the $\theta'$-Al$_2$Cu/$\alpha$-Al interface can either be 60° mixed or screw ones (Fig. 4a). Therefore, the interactions of these two types of dislocations with the $\theta'$-Al$_2$Cu precipitate are investigated separately in the simulation box (Fig. 4b).

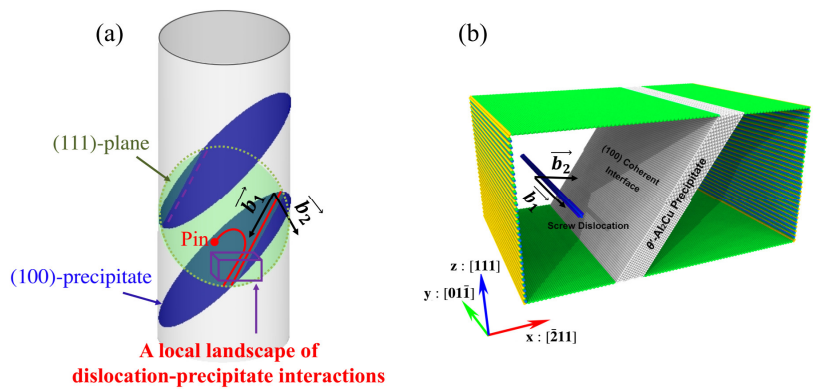

**Figure 4. Atomistic model of the MD simulation. (a)** Orientation relation between the (111) slip plane and the (100) precipitate. The dislocation deposited at the interface has a Burgers vector $\vec{b_1} = a/2\,[01\bar{1}]$ or $\vec{b_2} = a/2\,[10\bar{1}]$, corresponding to screw or 60° mixed dislocation respectively. **(b)** MD model representing a box inside the micro-pillar. The atoms of the $\theta'$-Al$_2$Cu precipitate are colored in gray, and the Al atoms in the perfect lattice are not shown. The initial dislocation is created near the surface (blue atoms).

In the configuration considered here (Fig. 4a), bypass of the penetrating plate-like precipitates from a classical Orowan mechanism is impossible, owing to the absence of space for the dislocation to bow-out. Therefore, the 60° mixed dislocations created in the MD simulation are obstructed when encountering the precipitate, forming a dislocation pile-up (Fig. 5a and b). The resulting stress concentration finally forces the leading dislocation to shear through the precipitate (Fig. 5c), leaving a highly disordered lattice along the shearing path within the precipitate (Fig. 5d). This simulation result is in excellent agreement with our TEM observations of the shearing steps and highly disordered lattice structure in the precipitate after deformation (see a relevant paper [11]). However, this mechanism is found to be irrelevant to the observed {100}-slips and will not be discussed in detail.

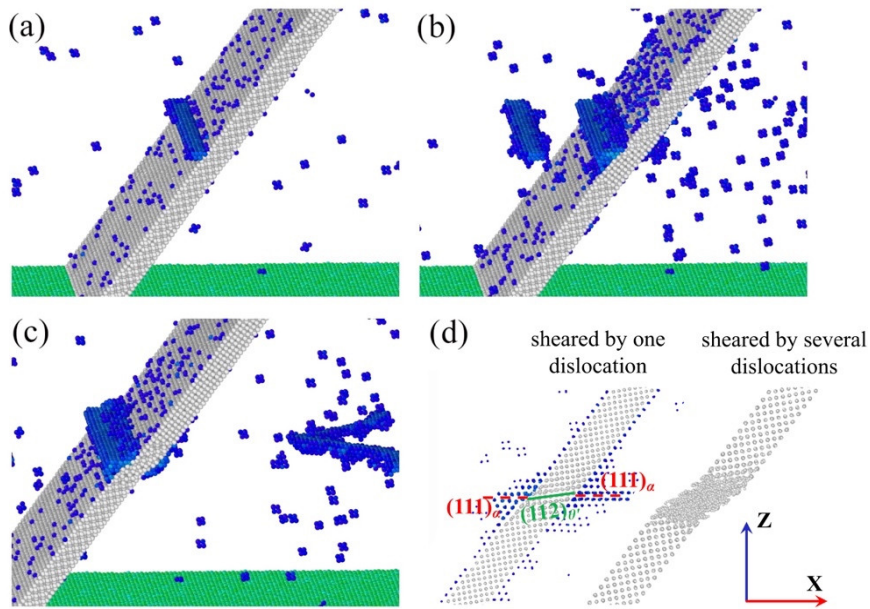

**Figure 5. Interactions between 60° mixed dislocations and a $\theta'$-Al$_2$Cu precipitate penetrating the entire section. (a)** A 60° mixed dislocation is blocked. **(b)** Dislocations pile-up against the interface. **(c)** The leading dislocation finally shears the precipitate. **(d)** The slip plane of $\theta'$-Al$_2$Cu precipitate is identified as $(112)_{\theta'}$ by analyzing the slip planes of a matrix dislocation before and after shearing the precipitate (left). After being sheared by several dislocations, a highly distorted lattice structure can be observed in the precipitate (right).

The situation is fundamentally different for screw dislocations. In this case, instead of precipitate shearing, our simulations reveal that screw dislocations can easily cross-slip from the (111) close-packed plane onto the precipitate/matrix interface under stress (Fig. 6a-c), and then move along the interface at room temperature (Fig. 6d). This cross-slip mechanism can lead to the {100}-slip traces observed in Fig. 2. However, in these MD simulations, stress values should be taken as rather rough estimates, owing to the high imposed strain rate (~ $10^8$ s$^{-1}$) and the numerical precision of the semi-empirical potential in representing the Peierls stress. As a result, the lattice resistance of interfacial slip cannot be quantified with great confidence. This leaves partly open the question of interfacial slip under experimentally observed stresses. Therefore, to confirm further the pertinence of the above deformation mechanism at experimentally observed stresses, a stress analysis will be presented in section 3.4.

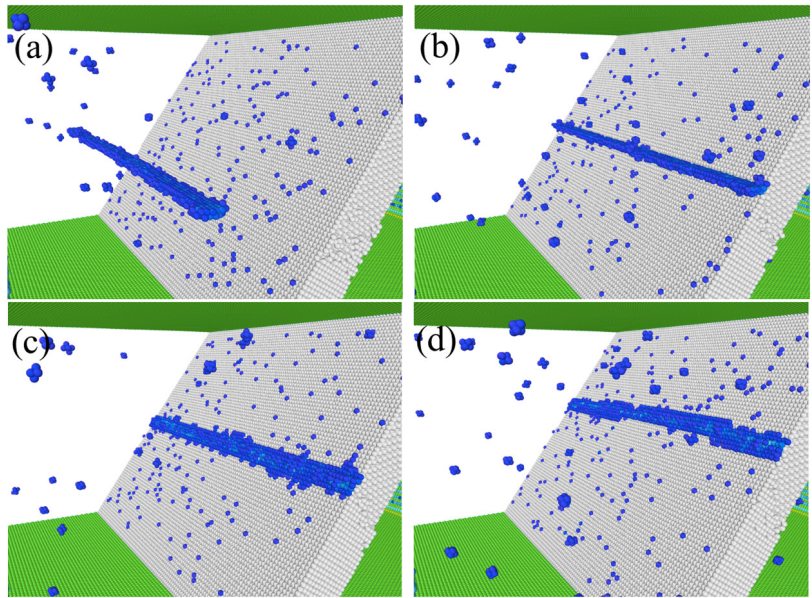

**Figure 6. Interactions between a screw dislocation and a $\theta'$-Al$_2$Cu precipitate penetrating the entire section.** A pure screw dislocation is created **(a)** and then moves towards the precipitate under shearing **(b)**. The screw dislocation cross-slips onto the (100) precipitate/matrix coherent interface **(c)** and moves along it **(d)**.

*3.3 Atomistic mechanisms of the interfacial slip*

The atomistic mechanisms determining the basic properties of the interfacial slip can be further obtained from MD simulations. The typical feature in the motion of interfacial dislocations (we call interfacial dislocations the dislocations cross-slipping from the {111} plane onto the interface) is shown on Fig. 7, which clearly supports a kink-pair mechanism: a kink-pair k1-k2 is thermally activated (Fig. 7b), and then extends along the straight dislocation during straining (Fig. 7c-e). As we applied periodic boundary conditions, the kinks k1 and k2 recombine with each other at the end, moving forward the dislocation for several atomic spacings (Fig. 7f). As mentioned above, in pure

Al, screw dislocations on {100} planes are unstable, with a strong tendency to dissociate on the intersecting close-packed planes [19]. As a consequence, dislocation motion proceeds through a kink-pair mechanism, with dissociated partials recombining from thermal activation back to {100} planes over a critical length, forming edge-kinks gliding on the non-compact plane [19, 20]. Instead, in our case, the interface stabilizes screw dislocations by preventing them from dissociating on the intersecting close-packed planes (Fig. 6), even after release of the external loading. Thus, the observed kink-pair mechanism should be understood as a result of a relatively large difference in lattice resistance between the edge kinks and the screw parts of the dislocation on the interface (see below).

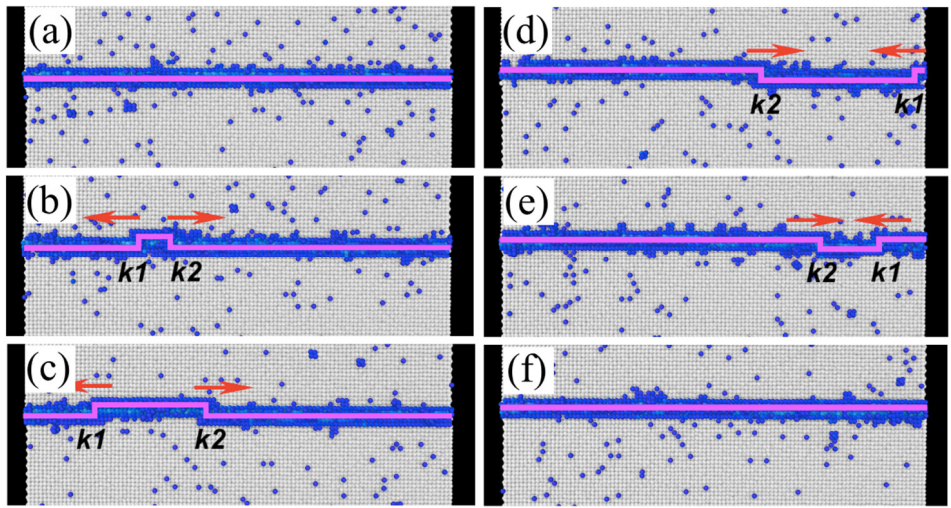

**Figure 7. Kink-pair mechanism of the screw dislocation moving on the coherent precipitate/matrix interface. (a)** A straight screw dislocation cross-slipped from (111)-plane to the (100)-interface. **(b)** Thermally activated nucleation of the kink-pair k1-k2. **(c), (d)** and **(e)** The edge kinks k1 and k2 move towards opposite directions. Note that boundary conditions are periodic in the direction of the dislocation line. **(f)** The screw dislocation advances several atomic spacings through the kink-pair mechanism.

To go further on the understanding of the interfacial slip, we explored the underlying atomistic mechanisms in more details by applying a numerical methodology within the P-N framework given in Ref. [31], which can provide both the dislocation core structure and the corresponding Peierls stress. In the classical P-N framework, the misfit region with inelastic displacement is assumed to be restricted to the glide plane of the dislocation. This lattice misfit generates short-range restoring forces between nearest neighbor atoms above and below the glide plane, as well as long-range elastic stresses [28]. The position of each atom on the glide plane can be calculated by considering the balance between the elastic stress and the atomic restoring force. More precisely, the atomic restoring force $F_b$ can be approximated from the generalized stacking fault (GSF) surface energy, $\gamma_{\text{GSF}}$:

$$F_b(S(x)) = -\frac{\partial \gamma_{\text{GSF}}(S(x))}{\partial S} \qquad (1)$$

where the disregistry $S(x)$ is the displacement of the crystal lattice above the glide plane with respect to the lower part at position $x$, and $x$ is the distance from the dislocation core in the direction of Burgers vector $\vec{b}$. The misfit density of the dislocation, $\rho_b$, can be defined as

$$\rho_b(x) = \frac{d}{dx} S(x) \qquad (2)$$

Then, the well-known P-N equation is obtained by writing the force balance between the elastic stress and the atomic restoring force:

$$\frac{k_b}{2\pi} \int_{-\infty}^{+\infty} \rho_b(x') \frac{1}{x - x'} dx' = -\frac{\partial \gamma_{\text{GSF}}(S(x))}{\partial S} \qquad (3)$$

For an isotropic medium, the energy coefficient of the dislocation, $k_b$, can be calculated from $k_b = G[(1-v)^{-1}\sin^2\theta + \cos^2\theta]$, where $\theta$ is the angle between the Burgers vector and the dislocation line. However, in the present case, we should take the elastic anisotropy and the effect of the interface into consideration. The $k_b$-value for a matrix dislocation is calculated by using the expression given in [40] for an anisotropic single crystal, while for an interfacial dislocation it can be obtained by following the procedure proposed for an anisotropic bicrystal [41]. The input elastic constants for Al are $C_{11}$ = 108 GPa, $C_{12}$ = 62 GPa and $C_{44}$ = 28.3 GPa, and $C_{11}$ = 190 GPa, $C_{12}$ = 80 GPa and $C_{44}$ = 90 GPa for $\theta'$-Al$_2$Cu precipitate [40]. Owing to their complexity, especially for the interfacial case, these calculations of $k_b$ are not re-detailed here. Following [40] and [41], $k_b$ is calculated to be 28.3 GPa and 38.9 GPa for screw and edge matrix dislocations respectively, and 35.3 GPa and 56.8 GPa for interfacial screw and edge dislocations respectively. To solve Eq. (3) numerically, the misfit function $S(x)$ is then approximated by the following expression [31]:

$$S(x) = \frac{b}{2} + \frac{b}{\pi}\sum_{i=1}^{N}\alpha_i \arctan\frac{x-x_i}{c_i} \tag{4}$$

where $\alpha_i$, $x_i$ and $c_i$ are adjustable constants, and $N$ is an integer. The normalization of $\int_{-\infty}^{+\infty}\rho_b(x)dx = b$ requires $\sum_{i=1}^{N}\alpha_i = 1$. Although Eq. (4) was originally proposed by assuming that the dislocation is composed of a number $N$ of elementary partial dislocations [42, 43], this physical interpretation has been discussed [44-46]. Instead, a more flexible mathematical interpretation can be proposed: substituting Eq. (4) into Eq. (2) gives $\rho_b(x)/b = \frac{1}{\pi}\sum_{i=1}^{N}\alpha_i\frac{c_i}{(x-x_i)^2+c_i^2}$, which is a sum of Lorentzian functions that can represent any bounded distribution (including $\rho_b(x)/b$ since $\int_{-\infty}^{+\infty}\rho_b(x)/b\,dx$ must be equal to 1 for a dislocation) with a sufficient number of $N$ [46]. Therefore, the problem simply reduces to search for parameters $\alpha_i$, $x_i$ and $c_i$ that can fit the real dislocation profile and fulfill the P-N equation (3). Note that in this case, the number $N$ does not correspond to the real number of partial dislocations, but only to the series of the approximation function of $\rho_b(x)$. The adjustable constants $\alpha_i$, $x_i$ and $c_i$ can be obtained from a least square minimization of the difference between the left and right sides of the P-N equation (3). $N = 3$ and $N = 6$ are found to be sufficient to provide a good fit for the matrix and interfacial dislocations respectively. With the fitted values of $\alpha_i$, $x_i$ and $c_i$, the misfit density of dislocation $\rho_b(x)$, which reflects the core structure of the dislocation, can then be obtained from Eq. (2) and Eq. (4).

The key parameter in solving the P-N equation (Eq. (3)) is the GSF energy $\gamma_{GSF}$, which can be calculated along the glide plane of the dislocation from ab-initio calculations [47, 48]. Thus, it is necessary to know the relative position between the glide plane and the interface. With this purpose, we examined the slip directions of each layer from MD simulations by viewing the lattice from the [100] direction (see Fig. 8b and c). The opposite slip directions found in Fig. 8c (see the arrows) indicate that the interfacial screw dislocations are formed from the relative glide between the layers of interfacial Al atoms and their neighboring Al atoms of the matrix (see the two columns of atoms marked by (c) in Fig. 8a). Therefore, the corresponding GSF energy curve should be calculated from ab-initio calculations by rigidly moving the part I over the part II of Fig. 8a in the slip direction $[01\bar{1}]$.

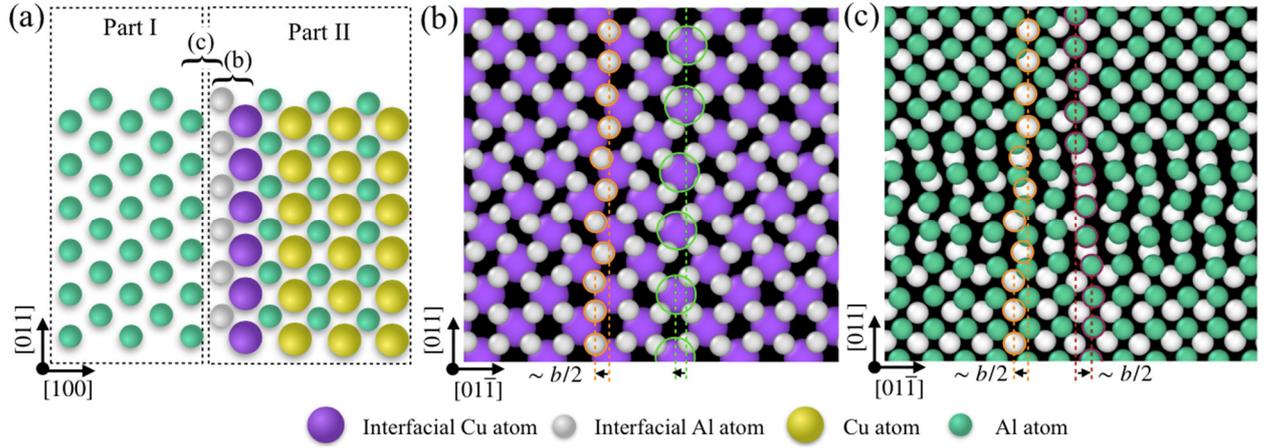

**Figure 8**. **Core structure of screw dislocation cross-slipped from (111) plane to the (100) interface.** **(a)** Lattice structure near the $\theta'$-Al$_2$Cu precipitate/Al matrix interface, where the interfacial Cu and Al atoms are colored in purple and white respectively. **(b)** and **(c)** Observations from the [100] direction for the atomic layers marked in (a). The atoms marked by circles show the slip direction across the screw dislocation in the middle of the image.

The GSF energy curve on the {100} slip plane identified above for Al-Cu alloys are presented in Fig. 9a, along with the corresponding curve on the {100} plane for pure Al. The significant difference in Fig. 9a indicates that the interfacial structure strongly modifies the GSF energy curve, hence the dislocation properties. From these GSF energy curves, the core structure of dislocations, reflected by the misfit density $\rho_b$, can be obtained by solving Eq. (3) from a least square minimization [31]. The single peak of the misfit density $\rho_b$ confirms the fact that the {100} dislocations in pure Al do not dissociate on their own slip plane [19, 49]. In this case, the core of the screw dislocations tends to spread on an intersecting close-packed plane to reach a lower-energy configuration [19]. In contrast, the two symmetrical peaks of the misfit density $\rho_b$ in Fig. 9c suggest that the interfacial dislocations tend to dissociate into partials strongly coupled on the coherent interface, with a plateau of misfit density in between. Besides the maximum peaks, secondary peaks in the misfit density as well as small fluctuations on the plateau (inset in Fig. 9c) can also be identified. They can be attributed to the sparser distribution of interfacial Cu atoms than the adjacent Al atoms in the matrix (see Fig. 8b), giving rise to strong stress fluctuations over the neighboring misfit atoms, and in turn leading to the fluctuations in the density profile [50]. In view of these features, i.e. dissociation on the {100} slip plane and the fluctuations in the misfit density, the core of interfacial dislocations is difficult to spread onto the intersecting close-packed planes, thus explaining the stabilization of screw dislocations on the interface observed in the MD simulations (Fig. 6).

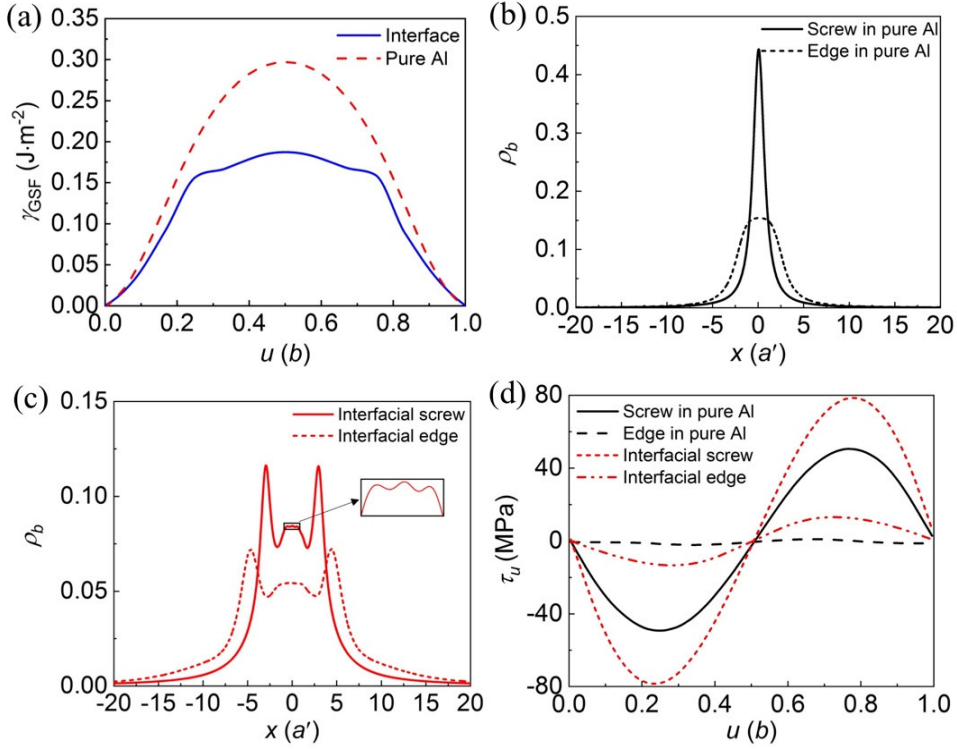

**Figure 9.** *Ab-initio* calculations of dislocations sliding along a $\langle 01\bar{1}\rangle$ direction on {100}-planes of the matrix or of the {100}-interface. **(a)** Generalized stacking fault (GSF) curves. The unite $b$ is the magnitude of the Burgers vector $a/2\langle 01\bar{1}\rangle$. **(b)** Misfit density of the dislocations plotted against the distance from dislocation center for dislocations on the {100}-plane of the matrix. **(c)** same as **(b)** for the {100}-interface. **(d)** Calculated gliding resistance, $\tau_u$, the maximum of which represents the Peierls stress.

In addition, the edge dislocations are generally wider compared with their screw counterparts (Fig. 9b and c), and thus should yield a much lower Peierls barrier along the slip direction. The Peierls stress, which is crucial to the stress analysis in section 3.4, is calculated as follows. If a dislocation moves a distance $u$ in the direction of Burgers vector, the misfit energy $E(u)$ can be written as a sum of GSF energies between pairs of atomic planes [31, 51]

$$E(u) = \sum_{m=-\infty}^{+\infty} \gamma_{GSF}\big(S(ma' - u)\big)a' \tag{5}$$

where $a' = a\sqrt{2}/4$ is the distance between two adjacent atomic planes, and $a$ is the lattice constant of Al (~ 0.404 nm). Note that we take the GSF energy on the exact slip plane of the interfacial dislocation into Eq. (5). The stress to overcome the barrier in the misfit energy is

$$\tau_u = \frac{1}{b}\frac{dE(u)}{du} \tag{6}$$

The maximum of $\tau_u$ represents the Peierls stress.

The Peierls stress for $a/2\langle 01\bar{1}\rangle\{100\}$ screw and edge dislocations in pure Al is estimated as ~ 50 MPa and ~ 1 MPa respectively (Fig. 9d). The fairly low Peierls stress for edge dislocation is compatible with the kink-pair mechanism proposed in pure Al by assuming a high mobility of the edge-kinks [19], and with real-time TEM observations showing easy motions of a single {100} edge dislocation in pure Al [52]. The interface increases the Peierls stress to ~ 80 MPa and ~ 15 MPa for the screw and edge dislocations respectively. This significant difference of Peierls stress, hence the dislocation mobility, between edge and screw dislocations serves as a base for the kink-pair mechanism, for both the {100}-slip in pure Al [19] and on the $\theta'$-Al$_2$Cu precipitate/$\alpha$-Al matrix interface (Fig. 7). With this interfacial Peierls stress for screw dislocation, a stress analysis is performed in the following section to check if the stress requirement for the mechanisms revealed by the MD simulations can be fulfilled in the micro-pillar.

*3.4 Stress analysis for the {100}-interfacial slip*

Under external stress, the dislocation can interact with the $\theta'$-Al$_2$Cu precipitate in three different manners: it can bypass the precipitate [6], shear the precipitate [10] or cross-slip along the precipitate/Al matrix coherent interface. Which mechanisms takes place largely depends on the relative magnitude of their slip resistance. In our samples, precipitate bypassing is impossible for geometrical reasons. It is therefore necessary to understand why the screw dislocation tends to cross-slip along the interface, rather than to shear the precipitate.

The slip system of $\theta'$-Al$_2$Cu precipitates has been identified as $a/2\langle 01\bar{1}\rangle\{211\}_{\theta'}$ from ab-inito calculations [9] as well as from our MD simulations (Fig. 5d). A local minima $\gamma_{sf}$ located at $u = a/2\langle 01\bar{1}\rangle$ can be found on the corresponding GSF curve extracted from Ref. [9] (Fig. 10a), indicating that the matrix dislocation will generate a stacking fault (SF) ribbon when shearing through the precipitate. The critical stress to generate this SF can be written as $\tau_{sf} = \gamma_{sf}/b = 1.06$ GPa [9], much larger than the Peierls stress along the interface (see above). On the other hand, it is possible that two $a/2\langle 01\bar{1}\rangle$ dislocations shear the $\theta'$-Al$_2$Cu precipitate as a pair, with a stable SF ribbon in between. In this case, the shear resistance should be determined by the Peierls barrier, rather than $\tau_{sf}$. Based on this consideration, we use the GSF curve in Fig. 10a to estimate the Peierls stress, following the same method as in section 3.3. The minimization result gives a dislocation profile showing two coupled $a/2\langle 01\bar{1}\rangle$ dislocations (Fig. 10b), and the corresponding Peierls stress is ~ 463 MPa and ~ 105 MPa for the screw and edge pairs respectively (Fig. 10c), larger than those of ~ 80 MPa and ~ 15 MPa for the interfacial counterparts (Fig. 9d). This stress comparison indicates that interfacial slip is easier than precipitate shearing for the $a/2\langle 01\bar{1}\rangle$ screw dislocations. For the 60° mixed type, precipitate shearing dominates as a result of a lack of cross-slip mechanism.

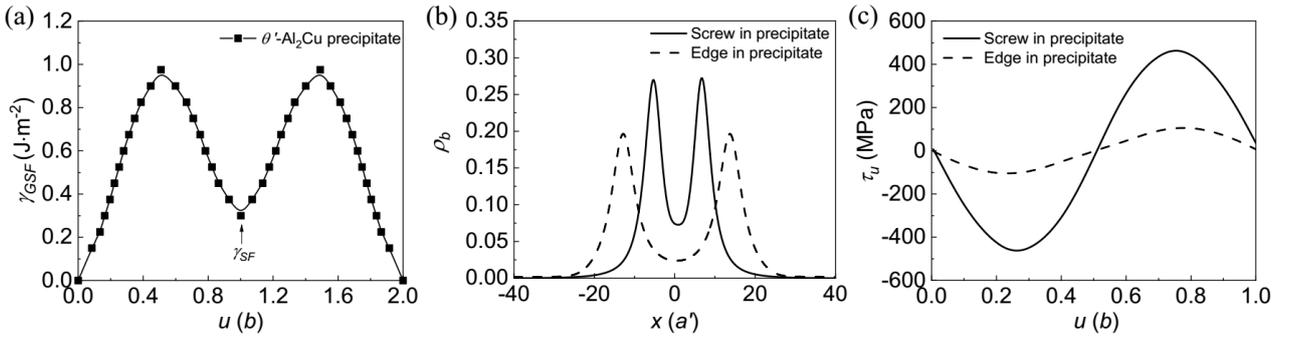

**Fig. 10. Ab-inito calculations of dislocations slipping along the $\langle 01\bar{1}\rangle$ direction on the {211} plane of $\theta'$-Al$_2$Cu precipitate. (a)** Generalized stacking fault energy taken from the color graph in Ref. [9]. The unite $b$ is the magnitude of the Burgers vector $a/2\langle 01\bar{1}\rangle$. **(b)** Misfit density of the dislocation plotted against the distance from dislocation center. The two peaks are related to two $a/2\langle 01\bar{1}\rangle$ dislocations coupled with each other. **(c)** Calculated gliding resistance, the maximum of which represents the Peierls stress.

In the following, we will further check whether the interfacial slip can be activated under the experimentally observed stress in micro-pillars. In the plastic regime, a leading dislocation would be initially blocked against the penetrating precipitate. If the external shear stress component on this leading dislocation exceeds the interfacial slip resistance, the cross-slip mechanism identified in the MD simulations (Fig. 6) can operate naturally. Otherwise, a dislocation pile-up will develop, concentrating the driving stress on the leading dislocation. Whether the interfacial slip can operate at experimentally observed stresses depends on the relative magnitude of this driving stress and the interfacial slip resistance. Note that in the MD simulations, we observed the operation of interfacial slip even without forming a dislocation pile-up (Fig. 6a). However, this cannot be directly transposed to experimental conditions, in view that the external stress and the interfacial lattice resistance in MD simulations are incomparable with those in micro-pillars. Once again, the MD simulations should be only considered as an investigation tool of the underlying atomistic mechanisms.

Given an external stress $\tau_{ex}$, the shear stress component at the head of a dislocation pile-up reads [40, 53, 54]

$$\tau_{shear} = (n-1)\tau_{ex} cos\varphi = \left(\frac{\xi \pi \tau_{ex} l_{sp}}{Gb} - 1\right)\tau_{ex} cos\varphi \tag{7}$$

where $n = \frac{\xi \pi \tau_{ex} l_{sp}}{Gb}$ is the number of dislocations piling up within a distance $l_{sp}$ between the pinning point of a single-arm dislocation source (SAS) and the precipitate [40]; $\tau_{ex}$ is the external shear stress on the plane of the pile-up, and simply obtained from the engineering stress at 2% here (~ 113 MPa, see Ref. [11]); $G$ is the shear modulus; $\varphi$ is the angle between the original {111}-slip plane and {100} plane of precipitate; $\xi = 1$ for screw dislocations [28]. The parameter $l_{sp}$ in Eq. (7), which represents the distance between the pinning point of a SAS and the precipitate, is likely associated with a large scatter, as the result of the random distribution of SAS within the micropillar, and is thus difficult to estimate precisely. For a sake of simplicity, we reasonably assume that the pinning points of the SAS are located, in average, midway between two penetrating precipitates, i.e., $l_{sp} = l_c/2$, where $l_c$ is the center-to-center distance between neighboring precipitates (calculated as ~ 678 nm from TEM measurements, following the method in Ref. [3]). By taking these parameters into Eq. (7), shear stress concentration at the head of a potential pile-up of screw dislocations can be obtained as ~ 1.2 GPa, *one order of magnitude* larger than the Peierls stress of interfacial slip (~ 80 MPa, see section 3.3). Note that this Peierls stress (at 0 K) represents an upper bound for the lattice resistance (at 298 K). This stress calculation confirms that the cross-slip mechanism observed in the MD simulations can indeed operate under experimental stresses observed during the compression of our micropillars, leading to {100}-slips along the precipitate-matrix interface.

It should be mentioned that interfacial slip occurs only when the dislocations cannot bypass the precipitates. In our case, this condition is fulfilled since the precipitates cross-over the entire cross-section, leaving no space for the dislocation to bow-out. In bulk materials, interfacial slip could be activated if the critical stress of bypassing is much larger than the lattice resistance of interfacial slip. This is not always the case for the deformation of Al-Cu alloy at room temperature. However, this mechanism might play an important role on dislocation recovery near precipitates when the accumulation of Orowan loops generate strong stress concentrations, hence on strain hardening [55]. On the other hand, in consideration of the associated thermally activated kink-pair mechanism, this interfacial slip should play an increasing role in bulk Al-Cu based alloys as elevating the temperature and/or decreasing the loading rate, i.e. on creep at elevated temperature. In this case, interfacial engineering, such as interfacial atom segregation or interfacial precipitation, will be crucial to hinder this interfacial slip, hence to enhance the creep resistance in these Al-Cu based alloys (see Ref. [56]). Our further work keeps going in this direction.

## 4. Conclusions

In this paper, we studied the effects of plate-like precipitates on the plasticity of Al-Cu micropillar, with a special sample size allowing the $\theta'$-Al$_2$Cu plate-like precipitates to cut the entire micropillar. We observed the unexpected presence of {100}-slip traces for the first time in Al alloys at room temperature. The atomistic mechanisms underlying these non-trivial {100}-slips are studied by performing MD simulations and theoretical analyses. The main conclusions are listed below.

(1) The {100}-slip operating along the coherent $\theta'$-Al$_2$Cu precipitate/$\alpha$-Al matrix interface is an intrinsic property in Al-Cu alloys, rather than an extrinsic behavior associated with Ga injection.

(2) The operation of single-arm dislocation sources can deposit screw dislocations or 60° mixed dislocations on the $\theta'$-precipitate/Al matrix interface. MD simulations indicate that 60° mixed dislocations can shear the $\theta'$-Al$_2$Cu precipitate when the precipitate cross-over the entire cross-section. This is consistent with our TEM characterization presented in our recent work focusing on the mechanical behaviors [11], but cannot explain the non-trivial {100}-slips.

(3) Instead, MD simulations and a stress analysis show that that deposited screw dislocations can cross-slip from {111} planes onto the $\theta'$-precipitate/Al matrix interface under the experimentally observed stresses, leading to {100}-slips observed experimentally at the position of the precipitates.

(4) The core of screw dislocations tends to spread on the {100} plane of the coherent $\theta'$-precipitate/Al matrix interface, in contrast with the unstable {100}-dislocations in pure Al dissociating in the intersecting {111}-planes. In addition, edge dislocations are wider compared with screw dislocations, thus yielding a much lower Peierls barrier in the slip direction. These two factors serve as a base for the kink-pair mechanism identified in MD simulations of interfacial slip.

## Acknowledgements

This work was supported by the National Natural Science Foundation of China (Grant Nos. 51621063, 51722104, 51625103, 51790482, 51761135031 and 51571157), the National Key Research and Development Program of China (2017YFA0700701, 2017YFB0702301), the 111 Project 2.0 of China (BP2018008), and China Postdoctoral Science Foundation (2019M653595). The financial support by the NSFC-ANR joint project SUMMIT (n° 17-CE08-0047, 51761135031) is sincerely acknowledged. We thank ShengWu Guo and YanHai Li for the TEM characterizations. Peng Zhang thanks the financial support from China Scholarship Council.